\documentclass[10pt, a4paper]{article}
\pdfoutput=1

 \usepackage{amsmath} 
\usepackage[sc]{mathpazo} 
\usepackage[english]{babel} 
\usepackage{hyperref} 
\usepackage[hang, small,labelfont=bf,up,textfont=it,up]{caption} 
\usepackage{graphicx} 
\usepackage{listings} 
\usepackage{array}
\usepackage[T1]{fontenc}
\usepackage{microtype} 
\usepackage[hmarginratio=1:1,top=32mm,columnsep=20pt, left=15mm]{geometry} 
\usepackage{multicol} 
\usepackage{multirow}
\usepackage{csquotes}
\usepackage{tabularx}
\usepackage{natbib}
\usepackage[usenames, dvipsnames]{color}
\usepackage{booktabs} 
\usepackage{float} 
\usepackage{lettrine} 
\usepackage{paralist} 
\usepackage{abstract} 
\usepackage{titlesec} 
\titleformat{\section}[block]{\large\scshape\centering}{\thesection.}{1em}{} 
\titleformat{\subsection}[block]{\large}{\thesubsection.}{1em}{} 
\usepackage{fancyhdr} 
\usepackage{enumitem}
\newlength\mylen
\setlist[itemize,1]{leftmargin=*,labelsep=-\mylen}
\usepackage{eurosym} 

\title{\vspace{-20mm}\fontsize{24pt}{10pt}\selectfont\textbf{Defining urban agglomerations to detect agglomeration economies}} \author{
\hspace{-8mm} \large \textsc{Cl\'{e}mentine Cottineau}$^{1}$, \large \textsc{ Olivier Finance}$^{2}$,\\
\hspace{-8mm}  \large \textsc{Erez Hatna}$^{1,3}$,  \large \textsc{Elsa Arcaute}$^{1}$,  \large \textsc{Michael Batty}$^{1}$\\[2mm]
\hspace{-8mm} \small $^1$ Centre for Advanced Spatial Analysis, University College London, W1N 6TR, UK. \\ 
\hspace{-8mm} \small $^2$ UMR 8504 G\'{e}ographie-cit\'{e}s, CNRS, and Universit\'{e} Paris 1 Panth\'{e}on-Sorbonne, 75005, France \\
\hspace{-8mm} \small $^3$ Center for Advanced Modeling in the Social, Behavioral and Health Sciences, Johns Hopkins University, Baltimore, 21209 USA\\
\hspace{-8mm} \small \href{mailto:c.cottineau@ucl.ac.uk}{c.cottineau@ucl.ac.uk}, \small \href{mailto:olivier.finance@univ-paris1.fr}{olivier.finance@univ-paris1.fr}, \small \href{mailto:ehatna1@jhmi.edu}{ehatna1@jhmi.edu}, \small \href{mailto:e.arcaute@ucl.ac.uk}{e.arcaute@ucl.ac.uk}, \small \href{mailto:m.batty@ucl.ac.uk}{m.batty@ucl.ac.uk}\\
\normalsize  
\vspace{-5mm}
}

\begin{document}

\maketitle 


\begin{abstract}
Agglomeration economies are a persistent subject of debate among economists and urban planners. Their definition turns on whether or not larger cities and regions are more efficient and more productive than smaller ones. We complement existing discussion on agglomeration economies and the urban wage premium here by providing a sensitivity analysis of estimated coefficients to different delineations of urban agglomeration as well as to different definitions of the economic measure that summarises the urban premium. This quantity can consist of total wages measured at the place of work, or of income registered at the place of residence. The chosen option influences the scaling behaviour of city size as well as the spatial distribution of the phenomenon at the city level. Spatial discrepancies between the distribution of jobs and the distribution of households at different economic levels makes city definitions crucial to the estimation of economic relations which vary with city size. We argue this point by regressing measures of income and wage over about five thousands different definitions of cities in France, based on our algorithmic aggregation of administrative spatial units at regular cutoffs which reflect density, population thresholds and commuting flows. We also go beyond aggregated observations of wages and income by searching for evidence of larger inequalities and economic segregation in the largest cities. This paper therefore considers the spatial and economic complexity of cities with respect to discussion about how we measure agglomeration economies. It provides a basis for reflection on alternative ways to model the processes which lead to observed variations, and this can provide insights for more comprehensive regional planning. \\


\end{abstract}


\begin{multicols}{2} 

\section{Introduction}

As complex systems, cities exhibit quantitative and qualitative change in composition as they grow in size. Economies of agglomeration are one of the most debated of such transformations. Research suggests that there is evidence of systematic variations in productivity levels across space, but the diversity of specifications used to estimate the magnitude of agglomeration economies makes it hard to compare studies and this leads to a wide array of differences in quantitative results. \\

In their review, \citet{rosenthal2004} find that city size tends to increase individual productivity by 3 to 8\%. In a meta-analysis, \citet{melo2009} looked at the parameters influencing this estimation in thirty-four studies where they found: 

\noindent\begin{itemize}
\item {\bf country specific} effects: stronger agglomeration effects are reported in studies of France and Italy, compared to the USA, and they are even weaker in China, Japan, Sweden and the UK.
\item {\bf industrial coverage} effects: Agglomeration economies of service activities are larger than those estimated for the economy as a whole.
\item {\bf controls and fixed} effects: Controlling for differences in human capital and skills at the individual level significantly lowers the estimation of city size effects on productivity. 
\item confounded relations from the type of {\bf urban measures} used (total population or density) and their {\bf spatial characteristics}. \item a {\bf publication bias} towards reporting a larger number of significant positive relations rather than negative or non-significant ones.
\end{itemize} 

Using six spatial definitions in France, \citet{briant2010} looked at scale and shape effects in the estimation of the wage premium. They concluded that their impact on the coefficients measured were less important than the effect of specifications and controls used in the regressions (the use of gross or net wages for example). We argue here that spatial effects are less trivial than shape and scale differences, and deserve to be studied extensively, in particular by distinguishing between urban and non-urban spaces (i.e. geographical units larger than cities at the regional scale or smaller than cities at the local scale). To this end, we ask three questions: \\

\begin{itemize}[label=--]
\item Are agglomeration economies specifically urban or do larger regions that include non-urban areas exhibit this behaviour too? 
\item Does the location of economic agents within cities matter when measuring agglomeration economies? 
\item Are larger cities more unequal?
\end{itemize}

These questions have not been studied much despite the fact that population cutoffs and spatial definitions affect the variation of {\it per capita} measures across city size \citep{fragkias2013, rybski2013, arcaute2015, cottineau2015}. In particular, the diversity of the urban landscape within cities is populated by different kinds of activities and different classes of population with different economic characteristics. We therefore expect that the way cities are defined matters in the estimation of agglomeration economies, the wage premium and internal inequalities.


\section{Hypotheses about cities, scales, economic output and inequality}

We start by reviewing the different economic mechanisms which might cause wealth, income and productivity to cluster in cities, in cities of larger sizes, and in neighbourhoods of particular affluence. We find that the mixed evidence from the literature is paralleled by a diversity in different theories as to the causes (section \ref{sec:cause}) and {\it loci} (section \ref{sec:geo}) of agglomeration economies. This leads us to articulate in section \ref{sec:why} our own hypotheses about why urban definition should be of importance in the estimation of agglomeration economies. 

\subsection{Mechanisms of agglomeration}
\label{sec:cause}

What we refer to here as a causal limit to agglomeration relates to the direction from which wage and income premia arise between cities and individuals as articulated in different economic theories. From the causes and reinforcing mechanisms of agglomeration economies within the framework of regional science, urban economics and the new economic geography, we find two groups of explanations. The first acknowledges a retroaction of cities on people (making them more productive, providing them with more knowledge, interactions and opportunities, etc.), whereas the second focuses on the micro foundations of aggregate measurements of the wage premium as well as of the spatial accumulation of wealth. \\

\citet{puga2010} summarises the first group into three classes of mechanisms. Larger cities and markets allow for increased: sharing (infrastructure and facilities), matching (demand and supply for products and skills) and learning (of new practices and technologies). These causal chains rely on the effect of city size and large-scale amenities with high fixed costs \citep{glaeser2001} to increase the output measured at the aggregate level. Another type of explanation involves the self sorting of the most productive and educated workers in the largest cities \citep{baumsnow2012}. Sorting describes the fact that economic agents of different educational attainments choose different locations, corresponding to different but equivalent bargains (typically: higher salary and higher rents in large cities or lower salary and lower rents in smaller cities), depending on their preferences for different sets of amenities. The distribution of jobs is assumed to follow the realisation of individual preferences and individual real income levels to be equivalent in cities of different sizes. However, the concentration of highly educated workers affects the productive composition of large cities and boosts the total income generated compared to smaller cities. \\

From an aggregate perspective, \citet{krugman1996} reviews potential mechanisms explaining the distribution of activity in space as an equilibrium between centripetal forces (natural and relative location advantages, market size, knowledge spillovers) and centrifugal forces (commuting costs, urban land rents, congestion and pollution) driving the location of firms and individuals as a consequence. In this framework, agglomeration economies occur where centripetal exceed centrifugal forces.\\

Consequently there seem to be two major ways of looking at higher wages and incomes in larger agglomerations: either urban concentrations produce higher productivity {\it per se}, or it is because highly productive workers and industries have similar location preferences. In the latter case, no real economic premium exists at the individual level in the largest cities as larger wages only compensate for larger rents, although \citet{eeckhout2014} suggest that this leads to larger cities being more unequal (due to complementarities in extreme skills). Similarly, either individual preferences or particular urban characteristics explain the variety of city sizes. These theoretical but undecided options suggest further empirical investigations, and add to the definitional sources of confusion: what are the most appropriate spatial limits of agglomerations and which economic quantity is most relevant to measure agglomeration effects?

\subsection{Defining agglomeration economies}
\label{sec:geo}

Two major sources of confusion (city definition and economic specifications) have hampered the empirical testing of results drawn from models that have originated in the New Economic Geography and the New Urban Economics. \\

\citet{martin1999} and \citet{osullivan2004} point to the unclear, undefined or usually over-simplified concepts of space and time which plague most theories of agglomeration economies. The extreme example of this simplification of space is the linear space that is assumed in the classic textbook by \citet{fujita2001}. Beyond this simplification, there does not appear to be any agreement about the scale at which agglomeration economies emerge. Most  authors refer to cities usually as homogeneous objects \citep{mera1973, sveikauskas1975, henderson1986, moomaw1988, henderson2003}, whereas some refer more broadly to industrial districts \cite{marshall1920}, dense urban spaces \citep{ciccone1996, ciccone2002} or to 'small regions' \citep{fingleton2006}. Some researchers have explicitly questioned the right scale at which these processes occur \citep{mori2015}. From a concentration analysis in the US performed at different geographical levels (from zip codes to states), \citet{rosenthal2001} conclude that different causes of agglomeration play out at each geographical level. \citet{arribasbel2015} examine the link between small employment districts and city-level measures of urban externalities and conclude that smaller units generate higher levels because cities are internally heterogeneous, with employment highly concentrated in their central parts and people of different income levels located in different areas. Finally despite limits on data, some theoretical urban arguments are empirically tested with regional data on exhaustive partitions, which seems problematic.\\

In this paper, we evaluate the hypothesis that the city is the relevant unit to observe agglomeration economies as well as internal inequalities. We do not take this statement for granted but explore and compare results for different geographies. Our exploration is motivated by a search for robustness but also because we think there are theoretical reasons why urban definitional criteria should play a role in the estimation of urban inequalities. \\

The second source of definitional confusion relates to what is measured empirically when referring to "economies", "rich populations" or "premium wages". Indeed, the lack of comparability between empirical studies from the literature emerges from the variety of economic outputs that relate to city size. Typically, the theoretical prediction which we want to verify with empirical data is the increase in added value per capita with city size (or agglomeration of population) \citep{mera1973, sveikauskas1975, ciccone1996}. However, added value is seldom measured at local levels of administrative geography, so wages are often taken as a proxy for productivity levels \citep{combes2011}. Moreover, taking nominal or real wages and controlling for education levels and/or housing prices have important impacts on empirical results \citep{moretti2004, briant2010}. We also observe a strong theoretical difference between measuring wage (or income) premium at the individual level of workers located in cities on the one hand \citep{briant2010, baumsnow2012, eeckhout2014}, and evaluating the effect of population size on the mass of income produced and its distribution in cities on the other \citep{glaeser2001, bettencourt2012, arcaute2015}. \\

Finally, when looking at inequality, a more classical indicator of wealth differences is income. This derives for some part from the revenue of labour (although with large variations across the income distribution of households), and for the rest from capital profits and welfare transfers \citep{piketty2013, atkinson2015}. These two specifications are thus not necessarily strongly correlated in space, as we know that the distribution of production and consumption at residence locations do not coincide \citep{davezies2008}. The uneven spatial distribution of income in cities is explained by models of segregation, but wage premia, income inequality, and segregation are rarely associated with city size even though the distribution of wealth is of paramount importance in urban processes of production and growth \citep{atkinson2015}. The key (and overarching) question of urban socioeconomics seems to be this: as cities get larger, do they get richer, but also more unequal and segregated \citep{batty2015, sarkar2015}? \\

Because all these processes are linked and localised in cities, we strongly expect their spatial delimitation to play a strong role in the empirical estimation of agglomeration economies and inequalities, as well as the obvious effects of econometric specification. We also expect these variations to tell a story about the social and morphological heterogeneity of urban space. We see four reasons why size effects on production and income inequality should vary with urban definitions.

\subsection{Why should urban definition matter to the estimation of agglomeration economies ?}
\label{sec:why}

\subparagraph{Because agglomeration economies are considered to be urban.}
If processes of learning, sharing, matching and sorting are enhanced by urban characteristics (such as the density of interactions), then the measure of agglomeration economies and scaling of economic variables needs to be performed on urban objects. A unique set of criteria to define cities is obviously illusory (as cities can be defined in many valid but non equivalent ways \citep{bretagnolle2001, parr2007}), but it is clear that the same estimation performed at an urban and a regional level are supposed to give different results. 

\subparagraph{Because the estimation of other parameters changes with city definition.} When studying the concentration of infrastructural and socio-economic attributes with city size, the choice of definitional criteria (density, flow and population thresholds to spatially delineate urban clusters) has proven to affect the value of scaling estimates \citep{fragkias2013, rybski2013, arcaute2015, cottineau2015}.

\subparagraph{Because cities are heterogeneously populated and used for activities.} Jobs and amenities are more concentrated than the resident population in cities \citep{terrier1990, glaeser2001, arribasbel2015}. This feature has increased since the 1950s because of the development of automobile mobility and the sprawl of residential population beyond the built-up area of cities. Therefore restrictive and extensive delineations do not capture the same activities within their urban boundaries and the resulting activity composition might be affected differently by size effects. Finally, in terms of social landscapes and income inequality, \citet{guilluy2013} and \citet{davezies2014} note for contemporary France, that different urban morphologies tend to coincide with different mixes of activities (productive and residential), but also with different population categories: recent immigrants and young professionals tend to reside in the densest parts of the largest metropolises, pockets of suburban wealth belong to their periphery, whereas rural areas and the outskirts of medium-sized and smaller cities concentrate deprived households. This implies that including some or all these heterogeneous spaces in the delineation of cities will modify the picture of urban wealth and inequality, and potentially its relation to city size. 

\subparagraph{Because geolocated estimations are subject to spatial biases in general.} There is a large body of literature that has focused on the aggregation biases in spatial measures associated with varying administrative boundaries. The Modifiable Area Unit Problem (MAUP) \citep{openshaw1979, openshaw1983} affects the measurement of concentration \citep{rosenthal2001} and segregation \citep{wong1999, wong2004, reardon2006} in general, so we must expect estimations of agglomeration economies and urban inequality to behave similarly and respond to the variations of shape and scale of the units defined as urban. The aim of this paper is to test this assumption and evaluate the significance of this variation on possible conclusions, using the data and methods exposed in the following section.


\section{Data and Methods}
\label{sec:method}
\begin{figure}[H]
 \begin{center}
  \caption{36546 Local Units (COM)}
  \label{fig:COM}
\includegraphics[width=0.3\textwidth]{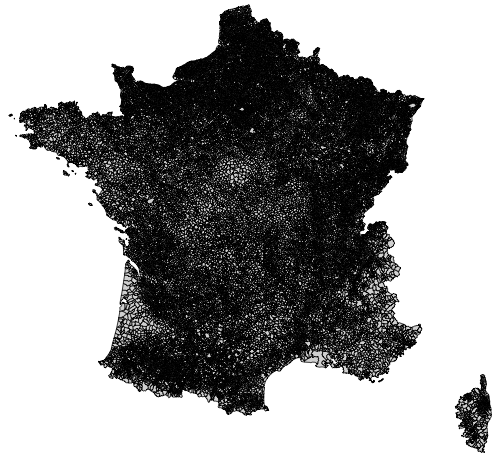}   
  \caption{2233 City Cores (UU)}
  \label{fig:UU}
\includegraphics[width=0.3\textwidth]{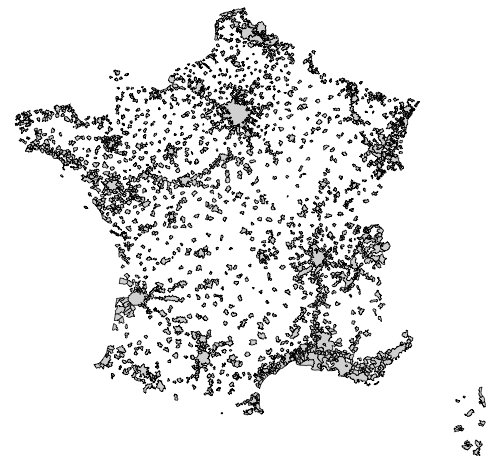}   
  \caption{771 Metropolitan Areas (AU)}
  \label{fig:AU}
\includegraphics[width=0.3\textwidth]{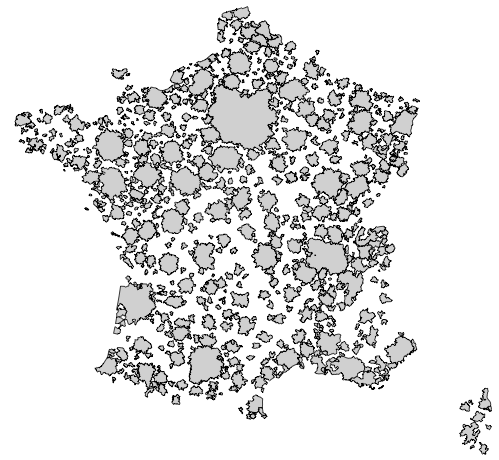}   
  \caption{96 NUTS 3 regions (DEP)}
  \label{fig:DEP}
\includegraphics[width=0.3\textwidth]{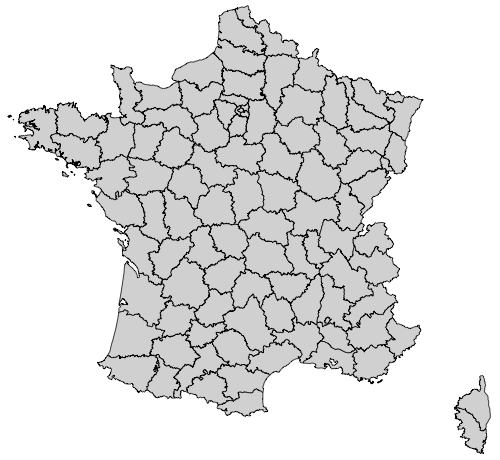}   
   \end{center}
\end{figure}

For this experiment, we use multiple definitions of cities applied to three sets of recent French data: 

\noindent\begin{itemize}
\item {\bf Population data} are gathered for 36546 local units from the 2011 French Population Census\footnote{source : \url{http://www.insee.fr/fr/themes/detail.asp?reg_id=99&ref_id=base-cc-evol-struct-pop-2011}} (cf. Figure \ref{fig:COM}) and aggregated into higher levels of geography (cf. Figures \ref{fig:UU} to \ref{fig:DEP}). 
\item  {\bf Income data} are provided by the taxation office. It gives for each local unit the sum of income declared by the 'fiscal households' (people who declare their income together) for the year 2011. The distributional income data refer to the number of households in each taxation bracket of declared income of 2011 and their aggregated income at the local unit level\footnote{source : \url{http://www2.impots.gouv.fr/documentation/statistiques/ircom2011/ir2011.htm}}. Unfortunately, the 8 taxation brackets do not correspond to regular quantiles of households. Their distribution is displayed in Table \ref{tab:income}. 
\end{itemize}

\begin{table}[H]
\begin{center}
\caption{Distribution of French households per income, 2011}
\label{tab:income}
\begin{tabular}{|c|c|c|c|c|}
\hline
Cat. & Income (k\euro) & N households & \% & of total* \\ \hline
B1 & $0 - 10$   & 6,380,662 & 24.0 & 18.1 \\
B2 & $10 - 12$   & 1,614,293 & 6.1 & 4.6  \\
B3 & $12 - 15$   & 2,616,818 & 9.8 & 7.4 \\
B4 & $15- 20$   & 4,279,602 & 16.1 & 12.2 \\
B5 & $20 - 30$   & 4,766,371 & 17.9 & 13.5 \\
B6 & $30 - 50$   & 4,323,569 & 16.3 & 12.3 \\
B7 & $50 - 100$   & 2,094,955 & 7.9 & 6.0 \\
B8 & $> 100$  \euro & 522,190 & 2.0 & 1.5\\ \hline
 & Total & 26,598,460 & 100 & 75.6  \\ \hline
\end{tabular}
\end{center}
\small *The number of households (35,178,358) is higher than the number of households for which we know the income bracket (26,598,460). We end up with 75.6\% of the distributional information.
\end{table}

For privacy reasons, categories with less than 10 households are absent from the data, leaving small local units with an incomplete description of their income distribution (1 out of 4 households are not accounted for in the distributional data, cf. Table \ref{tab:income}). This missing distributional data thus does not affect large (urban) units to the same extent as the smaller (rural) ones. \\

\noindent\begin{itemize}
\item
{\bf Wage data} comes from the public database CLAP ({\it Connaissance locale de l'appareil productif}\footnote{source : \url{http://www.insee.fr/fr/methodes/default.asp?page=definitions/clap.htm}}) which provides information about firms and their employees. We use data from 2008 for total wages and the total number of employees, which are gathered at the workplace (defined as the firm's location) and aggregated to the local scale of {\it communes}. In order to compare the distribution of firms according to their average wage levels with the distribution of income, we produced a partition into 8 classes that exhibits the same frequency as used in the income tax brackets (cf. Table \ref{tab:wages}). Because we had access to individual firms data, distributional information has a better spatial coverage than the income distribution. 
\end{itemize}

In order to compare the different specifications for the regressions performed here, we consider absolute numbers of firms, jobs and residents as measures of the size of a geographical unit, and the number of firms, jobs and residents per unit of urbanised surface (extracted from CORINE Landcover 2006 raster data\footnote{source : \url{http://www.statistiques.developpement-durable.gouv.fr/donnees-ligne/li/1825.html}}) as the density of a geographical unit.

\begin{table}[H]
\begin{center}
\caption{Distribution of firms and jobs per mean wages, 2008}
\label{tab:wages}
\begin{tabular}{|c|c|c|c|c|c|}
\hline
Cat. & Mean w (k\euro) & N firms  & \% & N jobs*   & \%  \\ \hline
F1 & $10 - 15.7$ & 277,421 & 24.0 & 1,235,134 & 11.6\\
F2 & $15.7 - 16.8$ & 70,187 & 6.1 & 327,983 & 3.1\\
F3 & $16.8 - 18.7$  & 113,776 & 9.8 & 625,324 & 5.9\\
F4 & $18.7 - 21.9$ & 186,071 & 16.1 & 1,192,853 & 11.1\\
F5 & $21.9 - 27.0$ & 207,234 & 17.9 &2,307,616 & 21.6\\
F6 & $27.0 - 36.8 $& 187,983 & 16.3 &3,344,014 & 31.3\\
F7 & $36.8 - 63.8$ & 91,085 & 7.9 &1,418,670 & 13.3 \\
F8 & $> 63.8$ & 22,705  & 2.0 & 222,805 & 2.1\\ \hline
 & Total & 1,156,462 & 100 &10,674,399 & 100 \\ \hline
\end{tabular}
\end{center}
\small *Jobs here correspond to full-time equivalents. Therefore, comparing the number of jobs (10,674,399) with the active population of France in 2008 (27,984,000) makes a satisfying approximation of the total productive activity in local units, as 2.1 million of them are unemployed, and around 20\% of the remaining 25.9 million work part-time. N.B. Establishments are the most relevant level to assign jobs, but wage data at this level are too weak. Therefore, we adopted the company level (as in most studies), as they provide a more exhaustive information. Consequently, in some cases, a job assigned to the company location may actually be performed in one of the company establishments in a different place. 
\end{table}

 Our method for defining cities (besides the official French delineations of urban cores {\it Unit\'{e}s Urbaines} and metropolitan areas {\it Aires Urbaines}) has been developed by \citet{arcaute2015} and applied to France by \citet{cottineau2015}. It consists of an algorithm for clustering local units ({\it communes} in France) into urban cores, based on a density cutoff: all contiguous local units of higher density than this threshold are aggregated. A second algorithm attaches functional peripheries to these urban cores, based on the percentage of commuters of local units working in the centres. A final criterion is used to select clusters: it is the minimum population of the considered cluster. The advantage of this method is that it can produce representations of the urban system for a variety of values for each criterion. Thirty different combinations are shown in Appendix \ref{fig:defs}.


\section{Empirical Results}
\label{sec:empirical}

Our empirical results are organised into three categories, following the three questions stated in introduction. Section \ref{sec:official} answers the question "are economies of agglomeration specifically urban" with mixed evidence. Section \ref{sec:system} reports variations of wealth across cities and other geographical units with respect to their size, trying to identify the definitions under which larger cities are richer. Finally, section \ref{sec:cities} investigates the question of inequality and its relation to city size.


\subsection{Are economies of agglomeration urban?}
\label{sec:official}

In this section we test whether or not agglomeration effects are characteristic of urban spaces. For this purpose we study two types of geographical boundary. The first type represents urban areas while the second includes non-urban regions as well. In the first and urban-only category, {\bf city cores} ({\it Unit{\'e}s Urbaines} or UU, cf. Figure \ref{fig:UU}) are defined based on the continuity of the built-up area ($< 200$ meters between buildings), and {\bf metropolitan areas} ({\it Aires Urbaines} or AU, cf. Figure \ref{fig:AU}) correspond to city cores of more than 1500 jobs to which are attached communes where more than 40\% of their working residents commute to the city core. In the second category, {\bf local units} ({\it Communes} or COM, cf. Figure \ref{fig:COM}) and {\bf NUTS 3 regions} ({\it D{\'e}partements} or DEP, cf. Figure \ref{fig:DEP}) provide an exhaustive coverage of the French territory, including rural areas, at respectively low and high scales. \\

These four delineations vary with respect to their number, population and spatial extension (cf. Appendix \ref{tab:4scales}). We use them as a benchmark to incorporate the effect of scale and 'urbanity' in different geographical delineations with economic specifications \citep{briant2010}.

Technically, we regress the total economic output (wages from firms and households' total income) of the spatial units on their total population and urban population density in the following equation
\begin{align}
\label{eq:scalinlog}
  log(Y_i)  = a_0  + \lambda \times log(U_i) + \epsilon_i
\end{align}
where {$Y_i$} represents the urban quantity (wages or income) of unit $i$ under enquiry, {$U_i$} is the urban referent (population or density), {$a_0$} is a time dependent normalisation constant, {$\lambda$} is the scaling coefficient and {$\epsilon_i$} is the estimated residual.\\

This regression is the log-transform of the following scaling equations specified for population \eqref{eq:scalingd} and density \eqref{eq:scalingensity}. 

  \begin{figure}[H]
 \begin{center}
  \caption{Variations of Scaling Exponents with respect to Geographical Scales and Economic Specifications}
  \label{fig:vizExponents}
  \includegraphics[width=0.47\textwidth]{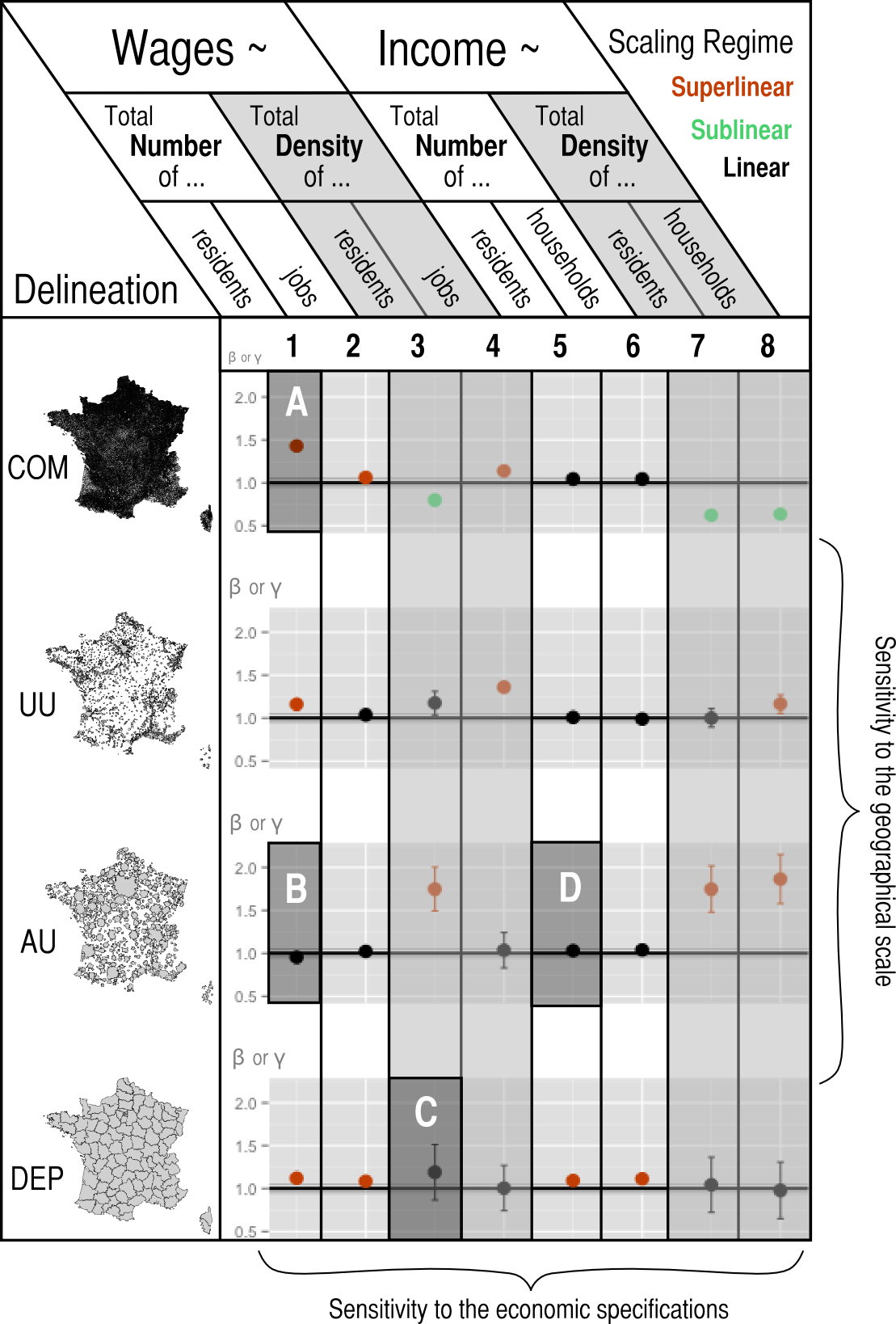}   
   \end{center}
\small From left to right on the x-axis: regression of total wages with total residents/total jobs ($\beta$), with the residential/job density ($\gamma$); regression of total income with total residents/households ($\beta$), with the residential/households density ($\gamma$). The dots represent estimated values, and the lines the extent of the confidence intervals.
\end{figure}

\begin{align}
\label{eq:scalingd}
  Y_i  = b \times P_i ^\beta 
\end{align}
where {\it b} represents a time dependent normalisation constant and {$\beta$} the scaling exponent of the variable with the size $P_i$ of unit $i$; and 

\begin{align}
\label{eq:scalingensity}
 Y_i  &= c \times D_i ^\gamma 
\end{align}
where {\it c} represents a time dependent normalisation constant and {$\gamma$} the scaling exponent of the variable with density $D_i$ of unit $i$.\\

We interpret the value of the exponents {$\beta$} and {$\gamma$} by reference to 1 (the linear regime), as is common practice in urban scaling \citep{pumain2004, bettencourt2007, arcaute2015}. In this case, an exponent equal to 1 represents the absence of economies of agglomeration. Exponents significantly greater than 1 (where the confidence interval does not intersect with the interval [0.95;1.05] for example in Figure \ref{fig:vizExponents}) indicate economies of agglomeration, that is rising average income or wages with size or density (the superlinear regime), whereas an estimation with $\beta$ or $\gamma$ significantly less than 1 (the sub linear regime) suggests diseconomies of agglomeration.\\

Figure \ref{fig:vizExponents} and present the results obtained with the four geographical delineations (rows) and eight economic specifications (columns): the OLS regression is performed on wages or income, using the absolute population or its density, and considering the unit's population as the resident population or the referent population (which is jobs for wages and households for income). Therefore, by looking at rows in Figure \ref{fig:vizExponents}, one can evaluate the sensitivity to economic specifications, whereas by looking at columns, the reader can see variations in estimations that depend only on the geographical delineation chosen (scale and 'urbanity'). The cells highlighted with letters correspond to estimations that can be compared with results from the literature.\\

We find that half of the estimations shown here (16 out of 32) give a linear relation between economic output and population (linear regime). Economies of agglomeration (superlinear) are found in 13 cases, and diseconomies (sublinear) in 3 cases, all at the most local level. Two of our estimations about cities in France are in agreement with similar estimations for UK cities where there are local economies of agglomeration in terms of wages with the resident population \citep{glaeser2001} (comparable in scale and specifications to Cell A) and no superlinear scaling of income in metropolitan areas \citep{arcaute2015} (Cell D). On the other hand, the absence of economies of agglomeration for wages with the total population that we find in metropolitan areas for France (Cell B) contradicts results usually produced for the USA \citep{sveikauskas1975, glaeser2001, bettencourt2012}. Similarly, the regression of wages against residential density in NUTS 3 regions shows too high a variability to be considered significant in the linear regime in France (Cell C) whereas \citet{ciccone2002} reports the existence of economies of agglomeration in European and American regions. \\

The referent population (i.e.: residents, households or jobs) plays a decisive role in estimation when considering the variation of wages with density. Indeed, jobs are spatially more concentrated than houses, and less ubiquitous. For all the other cases, this specification plays a negligible role. There appears to be more specification difference reported between regressions with the absolute population or with the density than between regressions of wages or of income. However, the two exponents $\beta$ and $\gamma$ are mathematically linked through the scaling of the urbanised area with population (cf. Appendix \ref{app.ScalingDensityTotal}), and therefore their different values provide less information about possible economic mechanisms than about the way cities of different sizes sprawl in extent. On the contrary, differences between specifications based wages or income represent the two sides of wealth: its creation and its redistribution, and how these are not spatially equivalent.\\

Finally, regarding the sensitivity of estimates to geographical specifications, we find the scale of analysis to be decisively important for the results obtained. We also find mixed evidence as to the 'urbanity' of economies of agglomeration. Looking at the scaling of wages with size for example (column 2), economies of agglomeration appear only for non-urban definitions! Indeed, partitions of the territory at the local (COM) and the regional (DEP) scales exhibit superlinear exponents (respectively 1.060 and 1.086, cf. Appendix \ref{tab:scalingUAUwages}), whereas city cores (UU) and metropolitan areas (AU) produce exponents closer to 1 (respectively 1.037 and 1.024). Counter-intuitively then, only larger regions and small places are richer, not cities. We think this results from the inclusion of rural areas in the non-urban definitions. Therefore, the qualitative difference between urban and rural places (or regions) create this quantitative result, rather than a change in economic behaviour for spatial units of larger population. When we consider density (column 4), estimates significantly over 1 are obtained for the most local scales. This confirms the explanations involving local processes within the labour market (sharing, learning, matching) rather than those of residential 'sorting', because this only holds for high work densities in the city centres (and not with high residential densities with high levels of amenities). The opposite is true for the scaling of income with population (column 6): larger regions only (DEP) exhibit an exponent greater than 1. Finally, the only 'urban' manifestation of economies of agglomeration is unexpected and this is the relative concentration of income in cities of high residential density (column 8). This process concerns the city scale as it plays in the opposite way for local places (COM).\\

To conclude, economies of agglomeration on the production side are urban insofar as 'urban' refers to the dense productive areas of central cities, and on the consumption side insofar as 'urban' refers to large-scale integrated entities between cities and metropolitan areas. However, insofar as economies of agglomeration should derive from the absolute concentration of people and jobs, then they appear to be non-urban and are observed better at the regional scale.


\subsection{Are larger cities richer?}
\label{sec:system}

To disentangle the mixed conclusions about 'urban' economies of agglomerations we used a multiplicity of delineations of cities that allow for the exploration of intermediary situations to that of official definitions and proceed to a sensitivity analysis to three definitional criteria of cities: minimum density, minimum commuting flows and minimum population size. 

 \begin{figure}[H]
 \begin{center}
  \caption{Examples of urban clusters definitions}
  \label{fig:ex}
\includegraphics[width=0.47\textwidth]{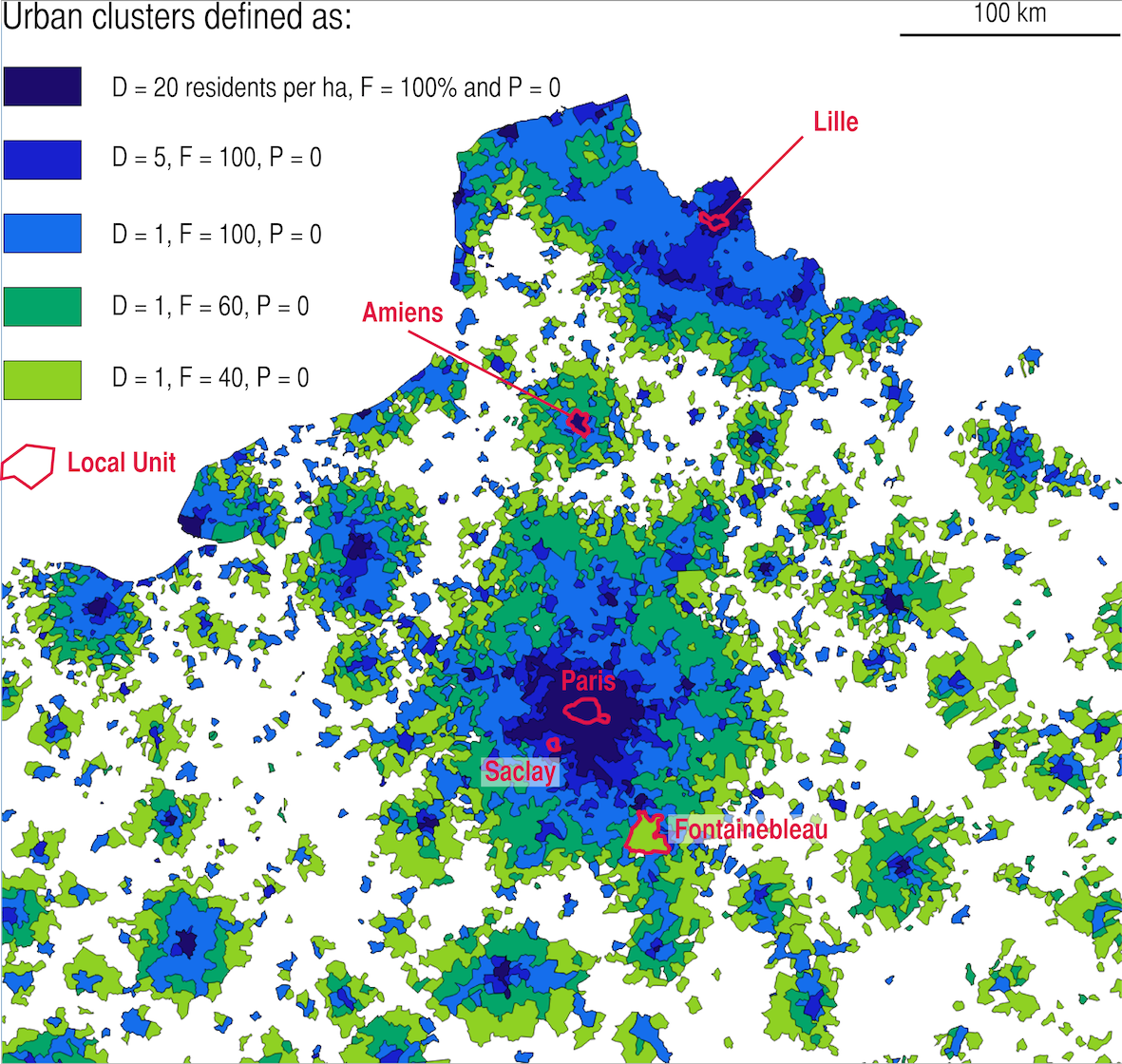}   
    \end{center}
\end{figure}

Using the algorithms developed by \citet{arcaute2015}, we combined 39 density cutoffs, 21 flow cutoffs and 6 population cutoffs to build 4914 representations of the French system of cities, from the most restrictive delineation (cf. Appendix \ref{fig:defs}, top right corner) of very dense centres with no functional periphery (close to local units or COM) to a very loose consideration of urban lifestyle that covers most of the French territory and thus close to a regional partition such as NUTS 3 or DEP (Figure \ref{fig:defs}, bottom left). 

 \begin{figure}[H]
 \begin{center}
  \caption{Synthetic urban clusters vs official definitions}
  \label{fig:corrDefs}
\includegraphics[width=0.28\textwidth]{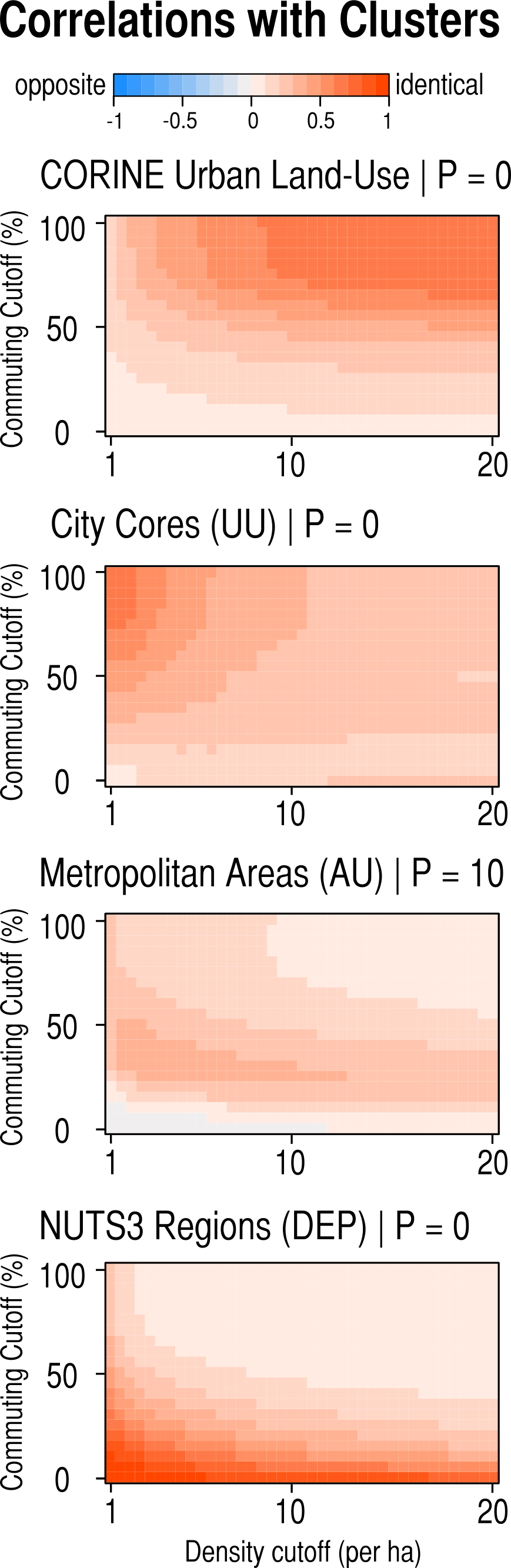}   
    \end{center}

  \end{figure}

The differences with the previous four definitions is that we can observe transitions of scaling behaviour for all three parameters that are easy to interpret, and thus control for the way people locate within cities. This answers our second research question. Figure \ref{fig:ex} illustrates the fact that each change in the definitional criteria affects the resulting urban clusters generated, and hence the measured levels of urban population, wage and income which we use. For example, when the density cutoff increases from 5 to 20 residents per ha, the technological pole of Saclay is excluded from the Parisian city core. When the density cutoff decreases from 5 to 1, the urban cluster of Lille embraces most of the North region. Finally, the large unit of Fontainebleau (which has high income but is covered in forest land-use) is not attached to the Paris metropolitan area over the 55\% flow cutoff. We finally see how the central and peripheral criteria affect in a somewhat different way dense agglomerations (like Lille for example) or suburban and smaller cities (such as Amiens). Figure \ref{fig:corrDefs} illustrates the correlation between the standard and synthetic definitions\footnote{The correlation analysis is performed at the level of the local units. If they belong to a cluster, for a given definition, they get the value 1, otherwise 0. The same process is done for UUs and AUs. At the regional level (DEP), all local units get the value 1, as the definition is a partition of the territory. Therefore a correlation of 1 means that all and only local units included in a city or a region are part of a cluster. A correlation of -1 on the other hand means that all and only local units included in a city or a region are not part of a cluster. For the CORINE raster dataset, the correlation is performed with the percentage of land use that is classified as 'urban' (111 + 112 in the 2006 land-use classification code).}.\\

Urban cluster definitions with high density cutoffs ($> 5$ residents per ha) and high flow cutoffs (i.e. composed mostly of central dense local units) represent best the urban land use at the local unit level as given by satellite raster data\footnote{From the CORINE LandCover 2006. A commune is labelled urban where the majority of its land-use is urban.}. The clusters that correspond best to standard City Cores (UU) have a minimum density cutoff ($< 4$ residents per ha) and a high flow cutoff ($> 70\%$). At similar levels of density, the cluster equivalent for Metropolitan Areas (AU) are characterised by a lower flow cutoff, as this definition includes the suburban spaces of commuting populations. Finally, the large clusters produced by definitions involving low density cutoffs and low flow cutoffs get closer to the complete coverage of France, and therefore to the spatial definition of NUTS 3 units.\\

\begin{figure}[H]
 \begin{center}
  \caption{A legend for the heatmaps of scaling exponents (Fig. \ref{fig:scalingTotalDefinitions}-\ref{fig:scalingseg})}
  \label{fig:readHM}
\includegraphics[width=0.4\textwidth]{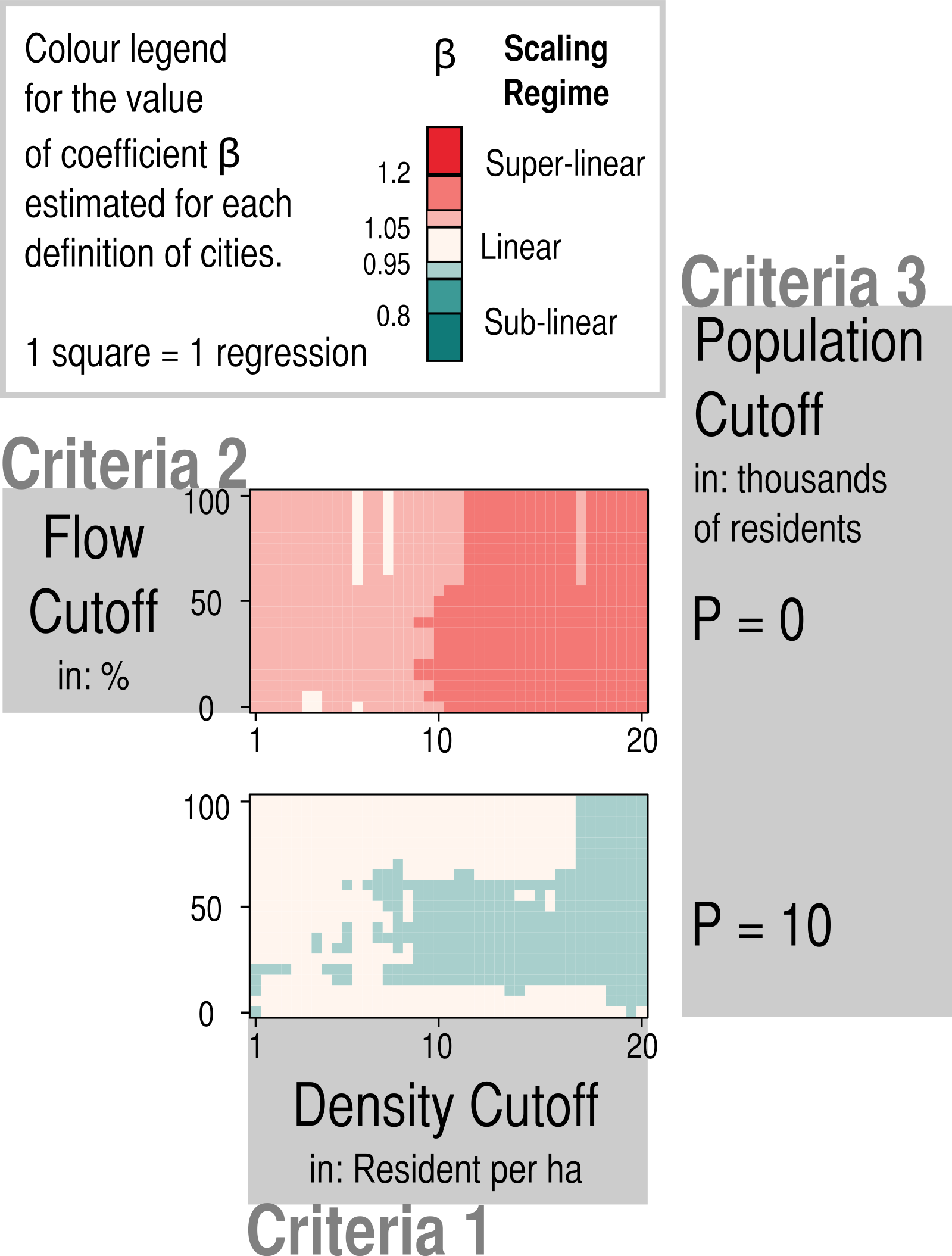}   
    \end{center}
\end{figure}

We regress total wages and income on the total and specific populations associated with wages and income, for each of the 4914 different definitions of cities. We relate the variations of the coefficients of the agglomeration economies with the definitional criteria of cities and the urban landscapes they correspond to by representing the value of coefficients by a colour and projecting this onto the definitional axes (density, flow and population cutoffs). Given the variety of definitional values that we take into account, we can trace the effect of these criteria on the estimation of the wage and income premium in French cities through heatmaps. Out of the 4914 regressions performed on 39 density cutoffs (criterion 1), 21 flow cutoffs (criterion 2) and 6 population cutoffs (criterion 3), we display only 3276 regressions in the following figures (4 heatmaps) by selecting 4 population cutoffs (no cutoff, 10, 20 and 50 thousand residents). The corresponding definitions are mapped in the same way as in Appendix \ref{fig:defs}, that is with cities as high density kernels in the top right corner, loose centres at the top left corner, and extensive cities at the bottom left corner. Each square represents the value of the estimated coefficient of the regression on all urban clusters defined by the three criteria (Figure \ref{fig:readHM}).\\

\begin{figure*}
 \begin{center}
  \caption{Variations of scaling estimations with city definitions}
  \label{fig:scalingTotalDefinitions}
\includegraphics[width=0.7\textwidth]{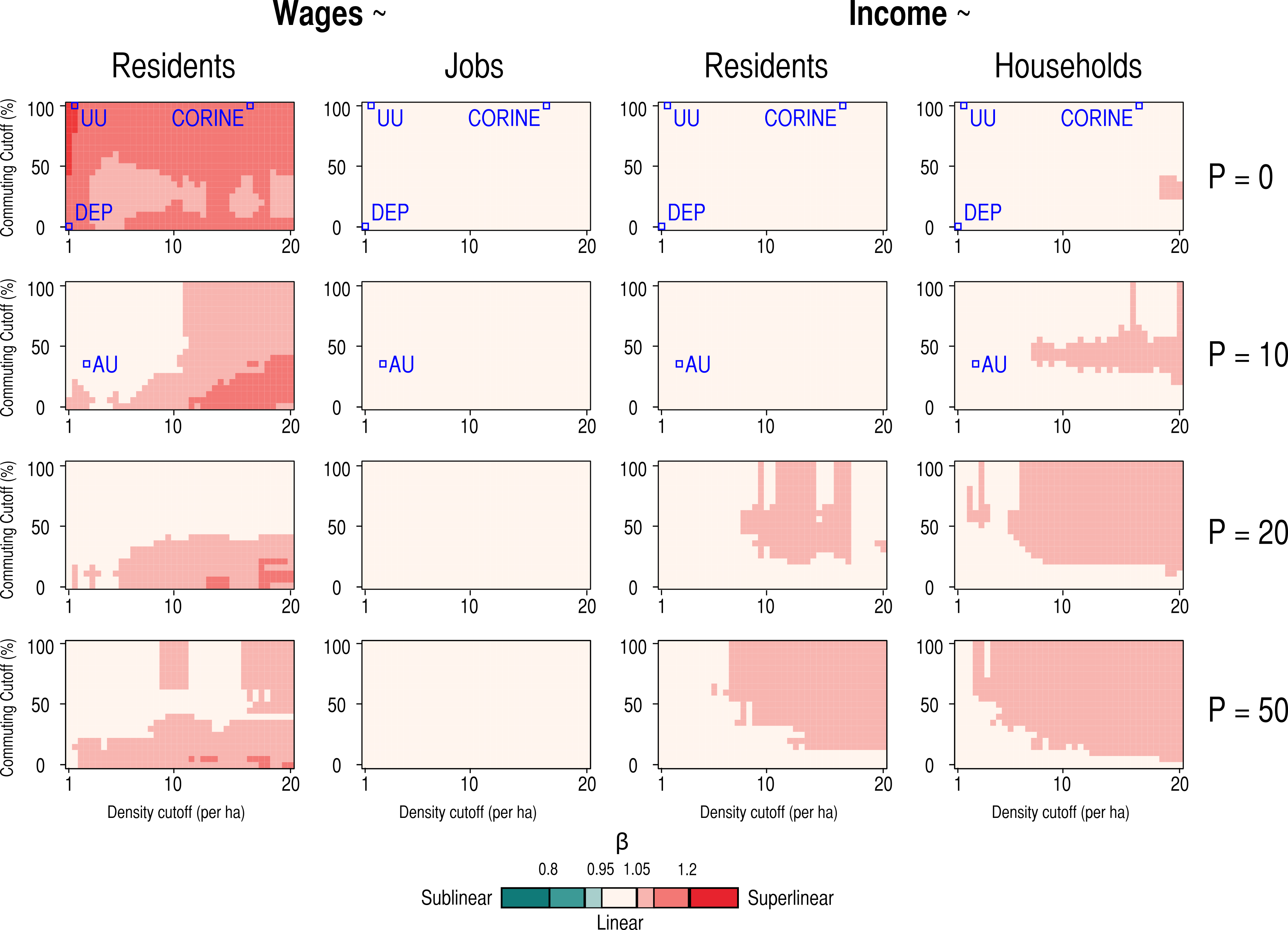} 
  \end{center}
\end{figure*}

When regressing the total wages and total income against total and specific populations (Figure \ref{fig:scalingTotalDefinitions}), we find mostly linear to superlinear relations. The projection of the standard definitions onto the closest systematic definitions helps understand the results obtained above. Indeed, Metropolitan Areas (AU) appeared at an urban scale at which no significant agglomeration economies are measured ($0.95 \leq \beta \leq 1.05$). This non-effect is also recorded for any definition when we regress the total wages by the number of workers at the workplace. For that matter, there are no values of parameters that enables us to observe agglomeration economies with this specification. Income begins to scale super-linearly with population at high density thresholds and for the upper part of the urban hierarchy only (where P $\geq 20$). This means that the relative concentration of income in large aggregates of population is a phenomenon that plays out at the level of large city centres only (high density cutoffs and high commuting flows), suggesting a change in urban quality when cities grow to become metropolises. Finally, the measure of agglomeration economies through wages (against total population) appears to be the most sensitive to city delineations: size effects become rare when we measure them on urban clusters with a minimum population of 10,000 residents (a reasonable minimum size for a city). This remains for clusters that do not correspond to any conceptual urban definitions (bottom right corner) or any population cutoffs. In this case, the coefficients measured are relatively high (superior to 1.05 and 1.1), matching observations made with local and city core definitions (cf. Figure \ref{fig:vizExponents}).\\

To conclude, larger cities tend to concentrate the jobs rather than higher wages. In terms of income, there are positive size effects in the most urban parts of the largest cities. Therefore, regarding our results for France, larger cities are not necessarily richer (this depends on the specification chosen) but the wealth produced in central cities 'circulates' \cite{davezies2008} spatially and ends up more concentrated as income per capita in the largest cities than it was as wages per job. The location of households and jobs within the cities thus definitely matters when it comes to estimating economies of agglomeration.


\subsection{Are larger cities more unequal?}
\label{sec:cities}

Thanks to the recent major contributions concerning inequality at national and international levels (for instance, see \citep{stiglitz2012, piketty2013, atkinson2015}), we know that the uneven distribution of wealth and income among the members of a society is not neutral as this affects social equity,  social reproduction, and economic growth. We also know that the spatial concentration of the rich \citep{pincon2010} and of the poor \citep{wilson1987, glaeser2009, bischoff2013} in neighbourhoods and cities can produce self-reinforcing effects of economic, health, crime and education inequalities \citep{blau1982, lynch1998, lobmayer2002, reardon2006, wilkinson2006, roscigno2006, manzo2010, aizer2014, dorling2014}. It has been reported elsewhere, generally in the USA, that larger cities are more segregated \citep{logan2011, bischoff2013} and more unequal \citep{glaeser2009, batty2015, sarkar2015}. Indeed, \citet{eeckhout2014} suggest that the better performance of large cities has to do with a productive complementarity between high skilled workers and low skilled workers, resulting in a larger inequality of skills and larger nominal wages (although they show a constant average of skills and real wages with city size). However, we are still missing a comprehensive understanding of size effects on segregation and inequality that integrate with the heterogeneous structure of cities.


\begin{figure*}
 \begin{center}
 \caption{Variations of the Gini index with city size and urban definition}
  \label{fig:scalinggini}
\includegraphics[width=0.7\textwidth]{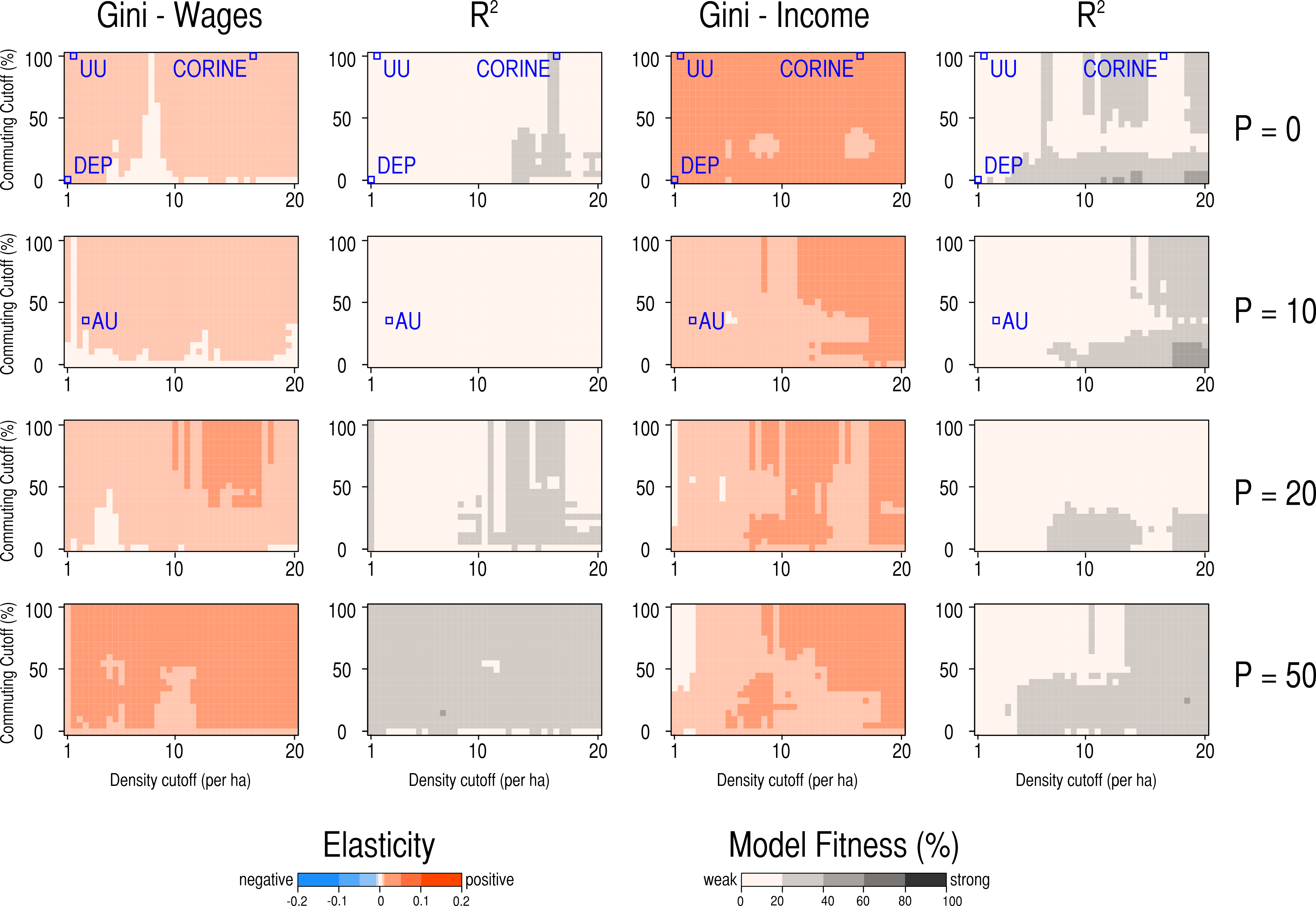} 
  \end{center}
\end{figure*}

\subparagraph{Inequality: } 
As an aggregate measure of inequality, the Gini coefficient has the merit of being synthetic, scale-independent and largely comparable between distributions based on different numbers and categories. Moreover, it correlates closely with other measures of inequality in cities \citep{glaeser2009} for which we have no information here. We have estimated Gini coefficients for wages and income from grouped data as in \cite{fuller1979}: 

\begin{align}
\label{eq:giniGroup}
G_{LowerBound} = \sum_{i = 1}^k (x_{i-1} y_i - x_i y_{i-1})
  \end{align}
\begin{align} \label{eq:giniGroup2} Gini = G_{LowerBound} + 0.5 g   \end{align}
where $k$ represents a group of income (or wages), $y_i$ is the cumulative proportion of income received by $x_i$, $x_i$ is the cumulative proportion of households earning $k$ or less (firms distributing average wages of $k$ or less) and $g =   \sum_{i = 1}^k  \frac{[(y_{i} - y_{i-1}) - ( r_{i-1} (x_{i} - x_{i-1})] [( r_{i} (x_{i} - x_{i-1}) - (y_{i} - y_{i-1})] }{r_i - r_{i-1}} $ with $r_i = \frac{z_i * X_k}{Y_k}$, where $Y_k$ is the total income, $X_k$ is the total number of households and $z_i$ is the lower bound of the bracket $i$.  \\

One value of the Gini index was computed for each city of each city definition, based on the number of households (or firms) of each income (or wage) category and their aggregate income (or total wages) in the city in question. The value of the Gini index was then regressed against the log of population for all cities {\it s} for a given city definition, as below:

\begin{align}
\label{eq:gini}
Gini_s = \alpha * log(Population_s) + b + \epsilon_s
\end{align}

Figure \ref{fig:scalinggini} shows the values of $\alpha$ as a function of the three definitional criteria of cities. A first conclusion is that size is not sufficient to predict the level of inequality in cities. Indeed, the statistical fitness of the regression models for all the city definitions are quite weak. Typically, less than 20\% of the variance of the Gini coefficients is 'explained' by the (logged) size of cities for definitions with low population cutoffs. For $P = 50$, most models 'explain' between 20 and 40\% of the variation in wage inequality, but only half of the variations of income inequality. That being said, this picture is however clear: all regressions clearly give a positive estimate. This coefficient belongs to the interval [0 ; 0.05] for any definition and economic specification. When $\alpha = 0.05$, inequality (the Gini Index) increases by $2^{0.05} = 3.5$ percentage points in a city twice as large. In other words, when there is a relation between size and inequality, it is always in the same direction: larger cities are slightly more unequal. The way one defines cities seems almost irrelevant with respect to this relation.


\subparagraph{Distribution: }

\begin{figure*}
 \begin{center}
  \caption{Variations of scaling estimations for distributional groups with city definitions}
  \label{fig:scalingIncomeDistribution}
  Number of Firms by mean {\bf Wages} categories vs. Firms population\\
  
\includegraphics[width=1\textwidth]{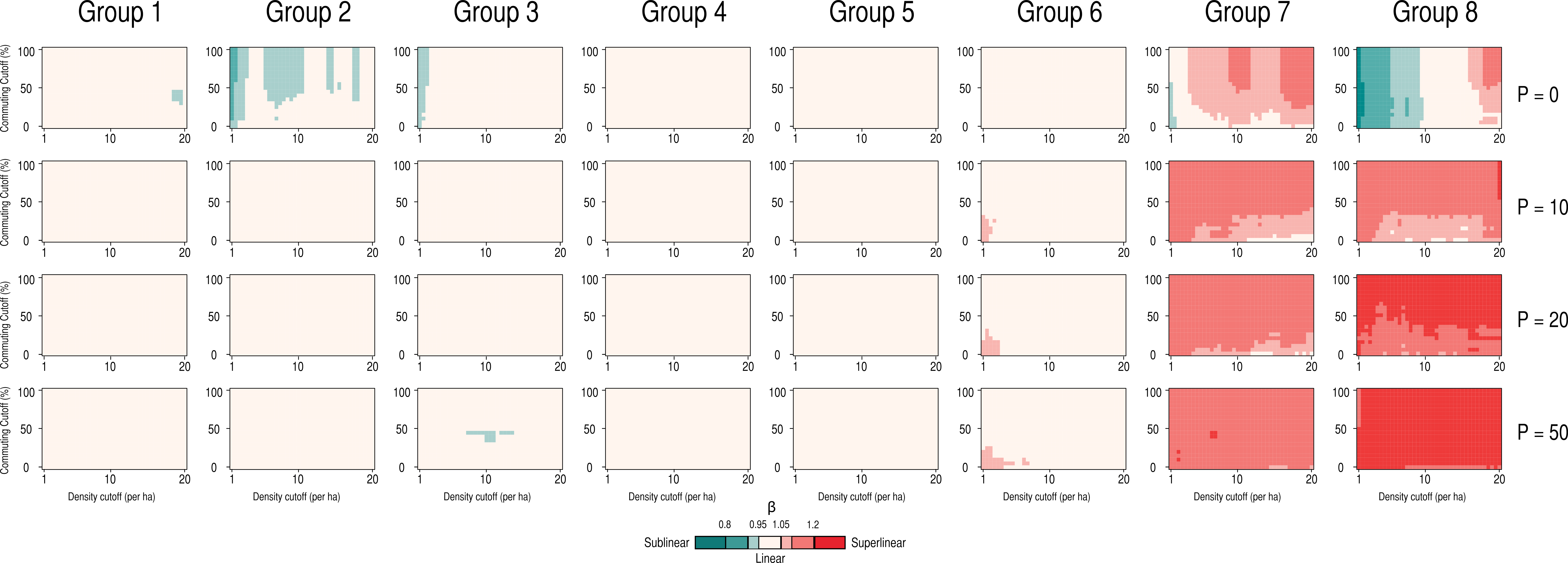}   \\

Number of Households by {\bf Income} categories vs. Households population\\

\includegraphics[width=1\textwidth]{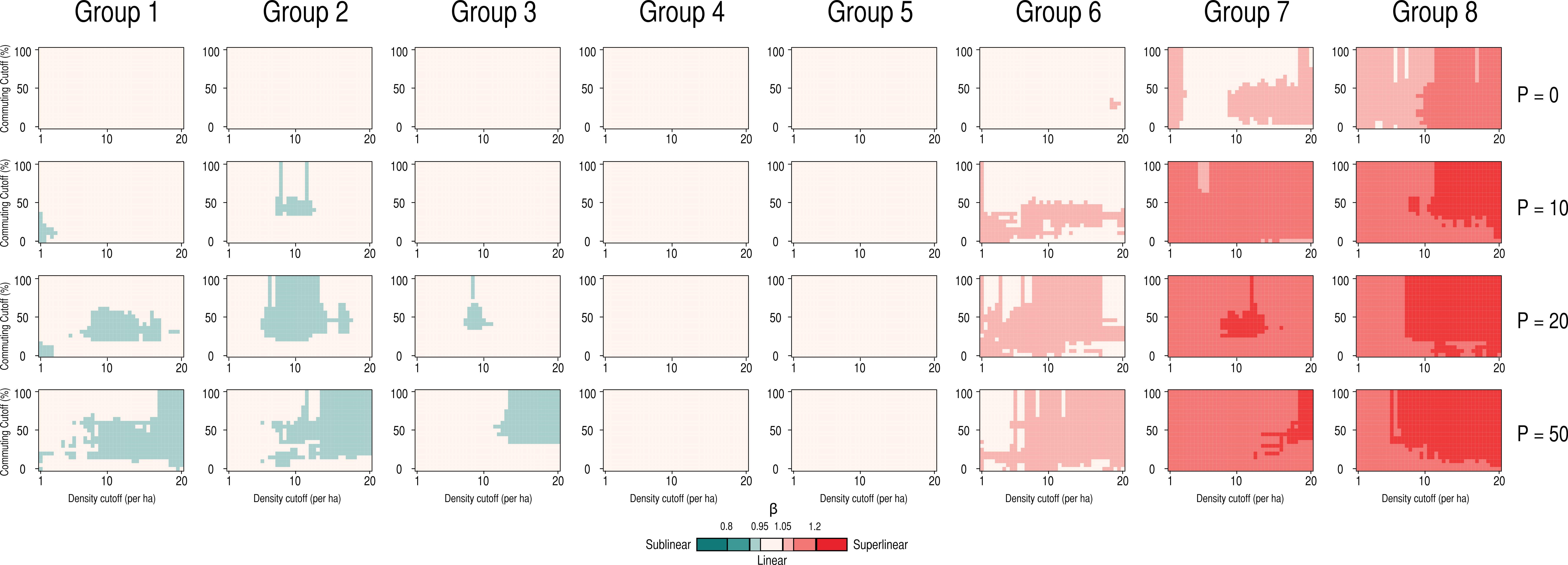}   \\

 Income and wage distribution \\
  \begin{tabular}{|c|c|c|c|c|c|c|c|c|}
\hline

  Group & 1&2&3&4&5&6&7&8 \\ \hline
Income (k\euro) & $0 - 10$ & $10 - 12$ & $12 - 15$  & $15 - 20$ & $20 - 30$& $30 - 50$& $50 - 100$& $> 100$   \\
Wages (k\euro) & $10 - 15.7$ &  $15.7 - 16.8$ & $16.8 - 18.7$ &$18.7 - 21.9$   & $21.9 - 27.0$ & $27.0 - 36.8$& $36.8 - 63.8$ & $> 63.8$ \\  \hline
\%& 24.0 & 6.1 & 9.8 & 16.1 & 17.9 & 16.3 & 7.9 & 2.0\\ \hline
  \end{tabular}
\end{center}
\end{figure*}

The Gini coefficient does not say if this higher inequality in large cities comes from a pyramidal society, if the rich are richer and/or if the poor are poorer in the largest cities. In order to disaggregate the study, we look at the income and wage distributions and investigate the scaling properties of the 8 wage groups of firms and of the 8 income groups of households. As stated in section \ref{sec:method}, the income partition is not ideal. The bounds are constrained by tax brackets and we have adapted them to represent the same share of firms in the case of wage data. Therefore, our income and wage groups are not significant by themselves but are at least comparable to one another. If larger cities are more unequal, we expect them to concentrate relatively more households in the low and/or high income groups (which would scale superlinearly), and firms of the low and/or high wage groups. By contrast, we expect middle-income and middle-wage groups to concentrate relatively in the smallest cities (and scale sublinearly). According to Figure \ref{fig:scalingIncomeDistribution}, the larger the city, the larger the number of high-income earners and the number of high wages firms. This observation is ordered by wage and income group, so that the richer the household and the firm, the more probable it is to find it in a large city. Therefore the hypothesis of a more pyramidal structure of the larger urban societies would be rejected, as we find a changing mix of firms and households with city size that looks more like a translation in the scale of average income and wages rather than a polarisation. \\

Comparing patterns of income and wages, we find a dual scaling behaviour for firms, as the potential wage premia are entirely due to the top 10\% of firms (classes 7 and 8)\footnote{and even more for the top 2\% of class 8}, which scale superlinearly. The 90\% remaining firms are distributed proportionately with the firm population in all cities, whatever the city definition. This pattern could induce more inequality between firms in terms of mean wages within the largest cities, but also a great inequality between cities of different size. The picture is similar for the income distribution, although in metropolitan areas of over 20,000 residents ($P = 20$), the lower income earners seem to scale sublinearly, whereas the top 25\% of households (class 6 to 8) scale super-linearly. This picture suggests that income inequality is a spatial phenomenon that is most striking at the national level between cities than within them, because households of similar income seem to concentrate in the same kinds of cities: small cities for the poorest, and large cities for the richest. \\

Finally, there are two exceptions and interesting variations of scaling of distributional groups with respect to city definitions: first increases in superlinear exponents for the top income earners in the most urban parts of cities (when density, commuting integration and population minimum increases - corner top right of the heatmaps), second the sublinear scaling of top wages firms in city definitions with low minimum density and no population cutoff. On looking at these regressions and their residuals, they usually include urban clusters with very few firms. We therefore attribute the sublinearity to the artifact of a few top average wage firms such as high-tech firms which locate in small cities of industrial valleys (like Heuliez in Cerizay, providing 30 jobs assembling electric cars) or to  very small firms (hotels, helicopter tours, etc.) with part time employees in highly touristic areas (Corsica, Saint-Tropez, Morzine, etc.). Such businesses are clearly exceptions rather than the rule.\\

A last interesting finding is that when we scale the total income and wages by group with total population, we find the same result, whereas when we scale the total income and wages by group with the population of the group, all relations are linear. Therefore the largest cities do not generate wage and income premia for the each distribution group \citep{sarkar2015} but rather concentrate the richest income earners and firms.


\begin{figure*}
 \begin{center}
 \caption{Variations of the Segregation index with city size and urban definition}
  \label{fig:scalingseg}
\includegraphics[width=0.7\textwidth]{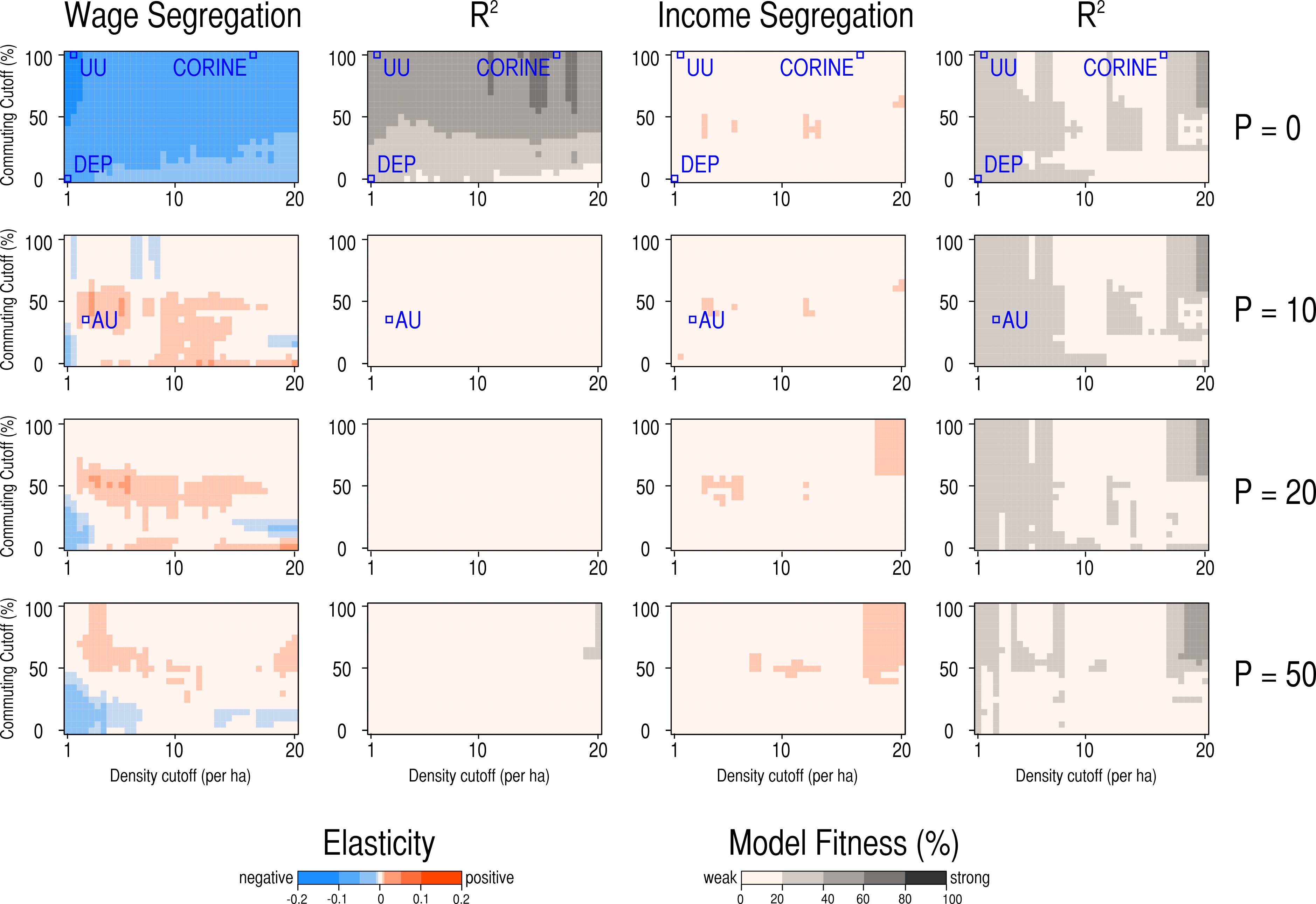}
  \end{center}
\end{figure*}

\subparagraph{Segregation: } A final way of looking at inequality in cities, along with the location of people and firms, is to look at the degree of the spatial segregation of economic groups within cities. Inequality and segregation are necessarily linked in the sense that if there were no economic inequality, there would be no economic segregation (or rather, it would be undefined). However, various spatial arrangements of economic agents are possible for the same level of inequality, depending on the growth of the housing market \citep{watson2006}, their total population \citep{bischoff2013} and/or their density profile \citep{spielman2014}. The social impacts of spatial clustering and segregation is known as the 'neighbourhood effects', and is usually positive in rich neighbourhoods, and negative in poor ones. Therefore it is worth investigating if the level of economic segregation varies with the different spaces in cities, as well as with respect to their size so that we might target urban policies more efficiently.\\

Income segregation is a relatively recent development in the segregation literature, and probably because of this, the indexes used to measure it are usually taken from those developed to measure residential racial segregation (a qualitative differentiation). The ordinal nature of income and wage groups however calls for a specific set of measures, such as the ones developed by \citet{reardon2009}. From this set, we chose the ordinal variation ratio index $R^O$ from equation \ref{R}, which we interpret as the "proportion of variation in a population that lies between, rather than within, organisational units" \citep[p. 150]{reardon2009}, in our case: the communes which compose the urban clusters. This measure does not depend on the overall inequality of each city, but only on the spatial distribution of groups present in the city.

\begin{align}
\label{R}
R^O = \sum_i{\frac{t_i}{T * v} (v - v_i)} 
\end{align}
 with: $i$ being a local unit that has a population $t_i$, and which composes a city cluster that has a population $T$, $v = \frac{4}{K - 1} \sum_j^{K-1}{c_j (1 - c_j)}$ and $v_i = \frac{4}{K - 1} \sum_j^{K-1}{c_{i,j} (1 - c_{i,j})}$. $K$ is the number of income (wage) groups $j$ and $c_j$ is the cumulative proportion of households (firms) which earn an income (distribute wages) of group $j$ or under.\\

One value of the segregation index was computed for each city of each city definition, based on the number of households (or firms) in each income (or wage) category in the local units composing the city in question. The value of the segregation index was then regressed against the log of population for all cities {\it s} as below:

\begin{align}
\label{eq:seg}
R^O_s = \alpha * log(Population_s) + b + \epsilon_s
\end{align}

Figure \ref{fig:scalingseg} shows the values of $\alpha$ as a function of our three definitional criteria of cities. A most unexpected result of this analysis is the decrease in wage segregation with size for the set of definitions with no population cutoff (the first heatmap on the left). Again, except for this case, the explanatory power of all the models is very low. For this one though, half of the value of the segregation index can be predicted by the number of firms in the city, no matter what the definition, and a larger number of firms leads to a lower level of wage segregation. The spatial segregation of average wages by firm decreases by up to 7 percentage points for a city twice as large. This means that there are larger variations of average wage between the firms of large cities than between the local units that compose them. In other words, the local mix of firms is more homogeneously representative of the whole in larger cities. This might derive from the very low number of firms in the smallest clusters, and the high possibility of being segregated: if two firms of two different wage group are in two different units, the segregation index equals 1. When small clusters are removed from the analysis, segregation of firms for the spatial equivalents of metropolitan areas increases with city size. \\

In terms of income, we find no conclusive size effects with respect to segregation. Except with very dense city core definitions (top right corners) and the equivalents for metropolitan areas, for which larger cities are slightly more segregated, we find elasticities close to 0 for all representations of the system of cities. Larger cities thus do not necessarily contain the greatest household segregation. This finding contradicts some of the evidence of increasing income  segregation with size in the US \citep{jargowsky1996, watson2009, glaeser2009, reardon2011, bischoff2013}. This could be explained by cultural differences between the two countries (large American cities being more segregated than large French cities), by attributes that are very specific to the definition of metropolitan areas (as we see a positive elasticity in the area that correspond best to the delineation of AU in France), by the different indexes used or by the fact that homogeneously rich and homogeneously poor cities are less segregated \citep{dabet2014}. The present data do no allow us to draw any conclusions about this matter.


\section{Conclusion}

There are a large diversity of theories and models concerning economies of agglomeration. A thorough evaluation of the sensitivity of empirical estimates to economic and geographic definitions can help us go beyond the mixed evidence reported (or censored in the case of negative results \citep{melo2009}) in the literature, and possibly refute, corroborate and choose between competing and complementary models. In this paper, we have reviewed the different causal mechanisms leading to positive externalities of urban agglomeration. We then hypothesised how different geographical specifications might affect the estimation of wage and income premia, before posing our three key questions:\\

{\bf Are agglomeration economies specifically urban or do larger regions that include non-urban areas exhibit this behaviour too?} Evidence from French cities and non-urban partitions reveal that economic specifications are crucial to answering this question. Economies of agglomeration with absolute size appear higher at a regional (i.e. not specifically urban) scale, but there seems to be a local productive advantage to local concentration of workers and at an urban scale, income concentration with density. These results back up micro-behavioural hypotheses of better learning, sharing and matching in a dense working environment. Sorting seems more appropriate to describe residential location choices and the search for amenities in denser cities.\\

{\bf Does the location of economic agents within cities matter when measuring agglomeration economies?} Obviously, it does. The concentration of jobs relative to the spatial distribution of population explains most of the discrepancy between results obtained with different economic specifications. What is more is that size effects affect different spaces of cities and regions differently: large and dense city cores concentrate more income and more high-income earners as they grow in size, whereas income is proportional to the number of households in metropolitan areas. \\

{\bf Are larger cities richer and more unequal? } We asked implicitly in the title to this paper: how to define urban agglomerations if you wish to detect agglomeration economies? Based on the results reported in the paper, we find that built-up areas correspond to the delineation that produces the most distinct economies of agglomeration in the most consensual understanding of the term (higher wage output with working population density). The absolute size effect on income also happens for these kinds of dense city cores. But this comes with higher inequality, slightly higher income segregation, and it ignores the fact that the people working in the dense part of the city, generating economic output by interacting closely during the day, tend to commute from other parts of the city which are less efficient (in terms of economic output) and consume more infrastructure as they grow in size \citep{cottineau2015}. \\

We think that it is important to take into account this spatial focus on income and wealth inequality, given its consequences for a city population's health, education, crime and equity, particularly in an era when inequality is increasing through  generations and over space, alongside the every greater polarisation of income in big cities.\\

\begin{center}
\large \bf \sc{Aknowledgements}\\
\end{center}

The authors acknowledge the funding the ERC Advanced Grants MECHANICITY (249393-ERC-2009-AdG) and GeoDiverCity (269826-ERC-2010-AdG) and thank Fabien Paulus for providing access to the CLAP database.
 

\end{multicols}

 \newpage
\LARGE \centering Appendix\\

\setcounter{equation}{0}
\setcounter{figure}{0}
\setcounter{table}{0}
\setcounter{section}{0}
\makeatletter
\renewcommand{\theequation}{A\arabic{equation}}
\renewcommand{\thetable}{A\arabic{table}}
\renewcommand{\thefigure}{A\arabic{figure}}

\section{Overview of pre-defined spatial delineations}
\label{tab:4scales}

\begin{table}[H]
\begin{center}
\begin{tabular}{|c|c|c|c|c|c|c|}
\hline
Spatial def. & N & Area (ha): mean & [min;max] & Pop.: mean & [min;max] & Orders of magnitude\\ \hline
COM & 36 546 & 2 553 & [1 ; 5 188] & 1 724 & [0 ; 2 249 975] & 6 \\
UU & 2 233 & 8 414 & [4 ; 1 238 155] & 21 760 & [605 ; 10 516 110] & 5\\
AU & 771 & 61 381 & [18 ; 4 877 170] & 69 295 & [2 169 ; 12 292 895]  & 4\\
DEP & 96 & 972 529 & [77 ; 2 769 560] & 656 983 & [77 156 ; 2 579 208] & 2 \\
\hline
\end{tabular}
\end{center}
\end{table}

\section{Mathematical link between Scaling with Total Population and Density}
\label{app.ScalingDensityTotal}

\small As the urban area usually scales with city size, we have a relation of the form:

\begin{align}
\label{eq:scalingp}
  A_i  = a \times P_i ^\alpha 
\end{align}
 {\small where $A_i$ represents the urbanised area of city $i$}\\

We can relate the scaling equation of the variable $Y$ with total population and scaling with density by expressing $\beta$ as a function of $\alpha$ \eqref{eq:scalingp} and $\gamma$ \eqref{eq:scalingensity}:

\begin{align*}
  Y_i  &= c \times D_i ^\gamma \\
    &= c \times (\frac{P_i}{A_i}) ^\gamma \\
    &= \frac{c}{a^\gamma}  \times P_i ^{ \gamma (1 - \alpha)} \\
    &= b' \times P_i ^{\beta'} 
    \end{align*}
 {\small with $b' = \frac{c}{a^\gamma}$ and $\beta' = \gamma (1 - \alpha)$}.\\

This gives

\begin{align}
\label{eq:diff}
\alpha > 1 \iff \gamma < \beta \\
\label{eq:2}
\alpha < 1 \iff \gamma > \beta \\
\alpha = 1 \iff \gamma = \beta 
    \end{align}

Given the scaling of urban surface with population in French cities ($\alpha_{UU}=0.856 [0.841;0.870], R^2=86.4\%$, and $\alpha_{AU}=0.888 [0.873;0.904], R^2=94.2\%$) we expect to find $\gamma > \beta$, as found in Table \ref{tab:scalingUAUwages} and Figure \ref{fig:vizExponents}, consistently with \eqref{eq:2}. \\

\section{Scaling estimations for official definitions}
\label{tab:scalingUAUwages}

\begin{table*}
Scaling of total wages with {\bf population}
\begin{center}
\begin{tabular}{lclc|lclclclclclcl}
\hline
Spatial def. & Reference & $\beta$ & CI (95\%)  & $R^2$ \%    & n \\ \hline
\multirow{2}{*}{COM*}      & Total Pop. & 1.428 & [1.420 ; 1.436] & 77.6 & 34969 \\
			 & Total Jobs & 1.060 & [1.059 ; 1.061] & 99.0 & 34992 \\ \hline
\multirow{2}{*}{UU}   & Total Pop.  & 1.160 & [1.136 ; 1.184] & 79.7 & 2233 \\
                          & Total Jobs & 1.037 & [1.034 ; 1.041] & 99.3 & 2233 \\\hline
\multirow{2}{*}{AU}      & Total Pop.  & 0.954 & [0.935 ; 0.973] & 92.9 & 771  \\
			 & Total Jobs & 1.024 & [1.019 ; 1.027] & 99.7 & 771 \\\hline
\multirow{2}{*}{DEP}      & Total Pop.  & 1.122 & [1.068 ; 1.177] & 94.7 & 96  \\
			 & Total Jobs & 1.086 & [1.070 ; 1.103] & 99.4 & 96 \\\hline
 \end{tabular}
\end{center}
Scaling of total wages with {\bf density}
\begin{center}
\begin{tabular}{lclclclclclcl}
\hline
Spatial def. & Reference & $\gamma$ & CI (95\%)  & $R^2$ \%    & n \\ \hline
	\multirow{2}{*}{COM}& Total Density & 0.798 & [0.769 ; 0.826] & 11.7 & 22775 \\
			& Job Density & 1.137 &[1.126 ; 1.149] & 62.0 & 22788  \\ \hline
\multirow{2}{*}{UU}  & Total Density & 1.175 & [1.037 ; 1.313] & 11.1 & 2232  \\
			 & Job Density & 1.360 & [1.303 ; 1.417] & 49.7 & 2232 \\ \hline
	\multirow{2}{*}{AU}& Total Density & 1.748 & [1.494 ; 2.003] & 19.1 & 771 \\
			 & Job Density & 1.037 & [0.831 ; 1.242] & 11.3 & 771  \\ \hline
\multirow{2}{*}{DEP} & Total Density & 1.191 & [0.869 ; 1.512] & 36.5 & 96 \\
			 & Job Density & 1.006 &[0.743 ; 1.269] & 38.0 & 96 \\ \hline
\end{tabular}
\end{center}
Scaling of total income with {\bf population}
\begin{center}
\begin{tabular}{lclc|lclclclclclcl}
\hline
Spatial def. & Reference & $\beta$ & CI (95\%)  & $R^2$ \%    & n \\ \hline
\multirow{2}{*}{COM}      & Total Pop. & 1.043 & [1.041 ; 1.044] & 97.6 & 36437 \\
			 & Households Pop. & 1.043 & [1.041 ; 1.045] & 97.1 & 36457 \\ \hline
\multirow{2}{*}{UU}   & Total Pop.  & 1.008 & [1.001 ; 1.015] & 97.1 & 2233 \\
                          & HH Pop. & 0.989 & [0.981 ; 0.997] & 96.4 & 2233 \\\hline
\multirow{2}{*}{AU}      & Total Pop.  & 1.031 & [1.024 ; 1.038] & 99.1 & 771  \\
			 & HH Pop. & 1.040 & [1.035 ; 1.046] & 99.4 & 771 \\\hline
\multirow{2}{*}{DEP}      & Total Pop.  & 1.095 & [1.065 ; 1.124] & 98.3 & 96  \\
			 & HH Pop. & 1.115 & [1.085 ; 1.144] & 98.4 & 96 \\\hline
 \end{tabular}
\end{center}
Scaling of total income with {\bf density}
\begin{center}
\begin{tabular}{lclclclclclcl}
\hline
Spatial def. & Reference & $\gamma$ & CI (95\%)  & $R^2$ \%    & n \\ \hline
\multirow{2}{*}{COM}      & Total Density. & 0.621 & [0.604 ; 0.639] & 17.5 & 22995 \\
			 & HH Density & 0.634 & [0.617 ; 0.652] & 18.0 & 23005 \\ \hline
\multirow{2}{*}{UU}   & Total Density  & 1.000 & [0.893 ; 1.108] & 13.0 & 2232 \\
                          & HH Density & 1.164 & [1.055 ; 1.272] & 16.5 & 2232 \\\hline
\multirow{2}{*}{AU}      & Total Density  & 1.748 & [1.479 ; 2.018] & 17.4 & 771  \\
			 & HH Density & 1.866 & [1.580 ; 2.153] & 17.5 & 771 \\\hline
\multirow{2}{*}{DEP}      & Total Density  & 1.045 & [0.723 ; 1.367] & 30.6 & 96  \\
			 & HH Density & 0.980 & [0.652 ; 1.308] & 27.2 & 96 \\\hline
 \end{tabular}
\end{center}
\small *COM stands for {\it communes}, the French local spatial units, UU for {\it Unit\'{e}s Urbaines} (the definition of City Cores), AU for {\it Aires Urbaines} (Metropolitan Areas) and DEP for the NUTS 3 level ({\it D{\'e}partements}).\\
\end{table*}

\newpage
\section{Multiple Representation of the French system of Cities}
      \label{fig:defs}
 \begin{figure}[H]
 \begin{center}
 \begin{minipage}[b]{\textwidth}
 \includegraphics[width=\textwidth]{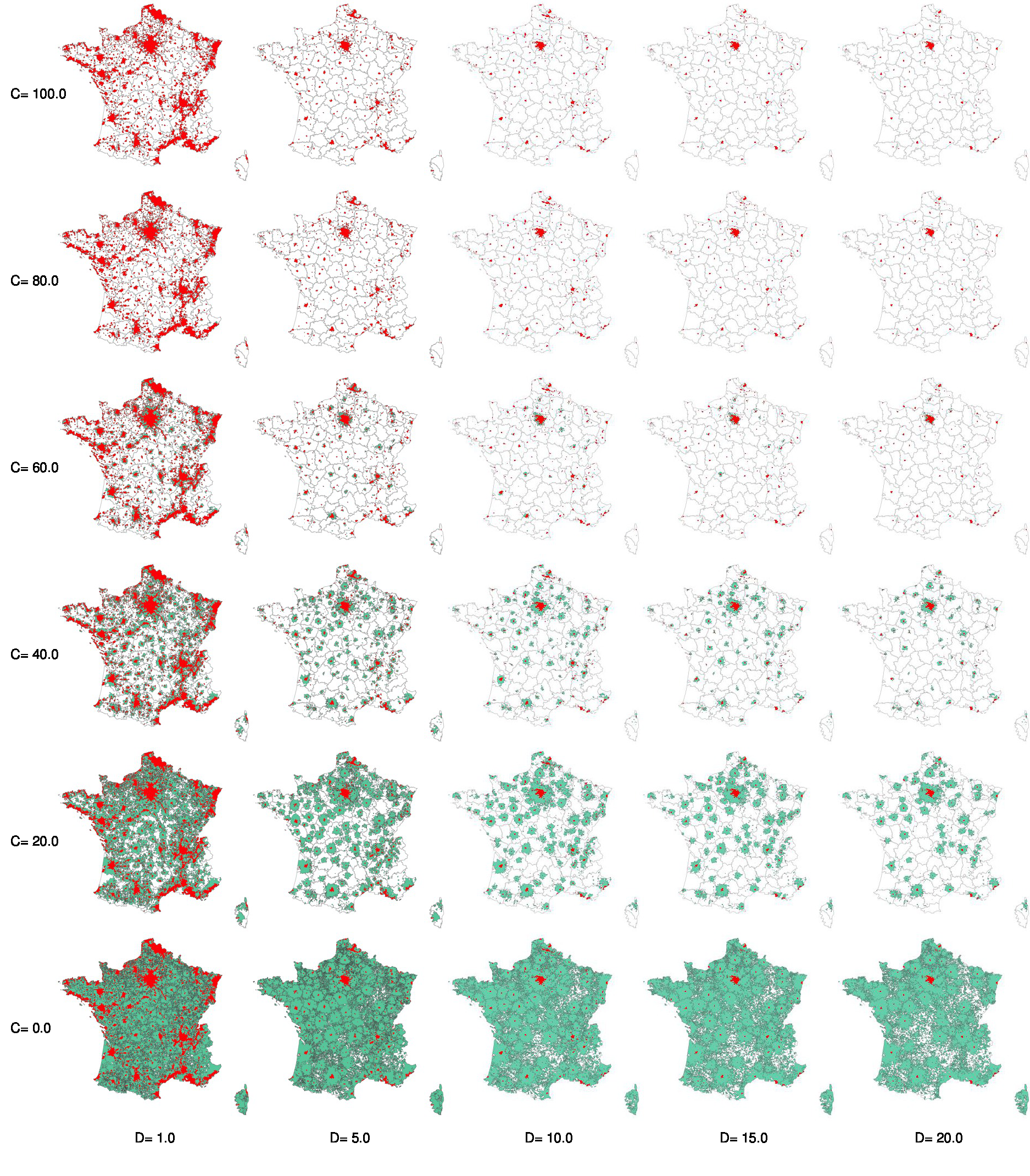}
P, the population minimum, is here set to 0. D is the minimum density of residents per ha to define urban centres (in red). C indicates the share of Commuters (in \%) living in the periphery and working in the density-based urban cluster (in green). Aggregation performed using the 2013 GeoFla geometry of communes. \url{https://www.data.gouv.fr/fr/datasets/geofla-communes/}
    \end{minipage}
    \end{center}
\end{figure}

\end{document}